\documentclass{emulateapj}

\def\ts     {\thinspace}
\def\kms    {\ts km\ts s$^{-1}$}

\def\aco    {$^{12}${\rm CO}($J$=1$\to$0)}
\def\bco    {$^{12}${\rm CO}($J$=2$\to$1)}
\def\cco    {$^{12}${\rm CO}($J$=3$\to$2)}
\def\dco    {$^{12}${\rm CO}($J$=4$\to$3)}
\def\eco    {$^{12}${\rm CO}($J$=5$\to$4)}
\def\fco    {$^{12}${\rm CO}($J$=6$\to$5)}

\def\ico    {$^{12}${\rm CO}($J$=9$\to$8)}
\def\jco    {$^{12}${\rm CO}($J$=10$\to$9)}
\def\kco    {$^{12}${\rm CO}($J$=11$\to$10)}

\slugcomment{draft version \today, accepted for publication in the Astrophysical Journal}

\shorttitle{CO(1--0) in z$>$4 QSOs}
\shortauthors{Riechers et al.}

\begin{document}

\title{CO(1--0) in z$\gtrsim$4 Quasar Host Galaxies: \\ No Evidence for Extended Molecular Gas Reservoirs}

\author{Dominik A. Riechers\altaffilmark{1}, Fabian Walter\altaffilmark{1}, Christopher L. Carilli\altaffilmark{2}, 
Kirsten K. Knudsen\altaffilmark{1}, K.~Y. Lo\altaffilmark{3}, \\ Dominic J. Benford\altaffilmark{4}, 
Johannes G. Staguhn\altaffilmark{4}, Todd R. Hunter\altaffilmark{5}, Frank Bertoldi\altaffilmark{6}, 
Christian Henkel\altaffilmark{7}, \\ Karl M. Menten\altaffilmark{7}, Axel Weiss\altaffilmark{7}, 
Min S. Yun\altaffilmark{8}, and Nick Z. Scoville\altaffilmark{9}}

\altaffiltext{1}{Max-Planck-Institut f\"ur Astronomie, K\"onigstuhl
  17, Heidelberg, D-69117, Germany}

\altaffiltext{2}{National Radio Astronomy Observatory, PO Box O,
  Socorro, NM 87801, USA}

\altaffiltext{3}{National Radio Astronomy Observatory, 520 Edgemont
  Road, Charlottesville, VA 22903-2475, USA}

\altaffiltext{4}{Laboratory for Observational Cosmology, Code 665,
  NASA Goddard Space Flight Center, Greenbelt, MD 20771, USA}

\altaffiltext{5}{Harvard-Smithsonian Center for Astrophysics, 60
  Garden Street, Cambridge, MA 01238, USA}

\altaffiltext{6}{Radioastronomisches Institut, Universit\"at Bonn, Auf
  dem H\"ugel 71, Bonn, D-53121, Germany}

\altaffiltext{7}{Max-Planck-Institut f\"ur Radioastronomie, Auf dem
  H\"ugel 69, Bonn, D-53121, Germany}

\altaffiltext{8}{Department of Astronomy, University of Massachusetts,
  710 North Pleasant Str., Amherst, MA 01003, USA}

\altaffiltext{9}{Astronomy Department, California Institute of
  Technology, Mail Code 105-24, 1200 East California Boulevard,
  Pasadena, CA 91125, USA}

\email{riechers@mpia.de}

\begin{abstract}
  We present \aco\ observations of the high--redshift quasi--stellar
  objects (QSOs) BR\,1202-0725 ($z=4.69$), PSS\,J2322+1944 ($z=4.12$),
  and APM\,08279+5255 ($z=3.91$) using the NRAO Green Bank Telescope
  (GBT) and the MPIfR Effelsberg 100\,m telescope.  We detect, for the
  first time, the CO ground--level transition in BR\,1202-0725.  For
  PSS\,J2322+1944 and APM\,08279+5255, our observations result in line
  fluxes that are consistent with previous NRAO Very Large Array (VLA)
  observations, but they reveal the full line profiles. We report a
  typical lensing--corrected velocity--integrated intrinsic \aco\ line
  luminosity of $L'_{\rm CO} = 5 \times 10^{10}\,$K \kms pc$^2$ and a
  typical total H$_2$ mass of $M({\rm H_2}) = 4 \times
  10^{10}$\,M$_{\odot}$ for the sources in our sample.  The CO/FIR
  luminosity ratios of these high--$z$ sources follow the same trend
  as seen for low--$z$ galaxies, leading to a combined solution of
  log$(L_{\rm FIR}) = (1.39 \pm 0.05) \times {\rm log} (L_{\rm
  CO})-1.76$.  It has previously been suggested that the molecular gas
  reservoirs in some quasar host galaxies may exhibit luminous,
  extended \aco\ components that are not observed in the higher--$J$
  CO transitions.  Utilizing the line profiles and the total
  intensities of our observations and large velocity gradient (LVG)
  models based on previous results for higher--$J$ CO transitions, we
  derive that emission from all CO transitions is described well by a
  single gas component where all molecular gas is concentrated in a
  compact nuclear region.  Thus, our observations and models show no
  indication of a luminous extended, low surface brightness molecular
  gas component in any of the high--redshift QSOs in our sample.  If
  such extended components exist, their contribution to the overall
  luminosity is limited to at most 30\%.
\end{abstract}


\section{Introduction}

Understanding when and how galaxies form is one of the primary
objectives in both observational and theoretical astrophysics. The
mere fact that active galaxies, such as radio galaxies, QSOs, and
emission line galaxies are observed up to redshifts of $z = 6.6$
(e.g., Fan et al.\ 2001; Rhoads et al.\ 2001; Hu et al.\ 2002; Kodaira
et al.\ 2003; Taniguchi et al.\ 2005), less than one Giga--year after
recombination, implies that galactic scale ($\sim$10\,kpc),
gravitationally bound structures exist at this early epoch. The study
of the masses and dynamical state of these young systems serves as a
direct constraint to the models describing the growth of large scale
structures since the epoch of recombination.

Studies of the molecular and dusty interstellar medium (ISM) in these
galaxies are of fundamental importance, since it is this medium out of
which stars form; accurate determination of the molecular gas mass may
therefore serve as an indicator of the evolutionary state of a galaxy.
The detection of carbon monoxide (CO) is also a strong confirmation
that star formation is going on in some of the highest redshift
systems. In fact, the combination of molecular gas and dust detections
with large far--infrared (FIR) luminosities provides the strongest
evidence that a significant fraction of high--$z$ galaxies is
undergoing starbursts at prodigious rates ($>10^3\,$M$_\odot {\rm
  yr}^{-1}$), consistent with the formation of a large elliptical
galaxy on a dynamical timescale of $\sim$10$^7$--10$^8$\,yr.

Over the past decade, more than 30 galaxies at $z>2$ have been
detected in CO emission (Solomon \& Vanden Bout 2005) out to a
redshift of $z=6.42$ (Walter et al.\ 2003, 2004; Bertoldi et al.\
2003), confirming the presence of intense starbursts in numerous
high--$z$ galaxies. As most of these observations were obtained using
millimeter interferometers, these detections were typically achieved
by observing high--$J$ CO($J \to J-1$) transitions ($J \geq 3$).
Although these high--$J$ lines exhibit in general higher peak flux
densities than the ground--state ($J=1$) transition, it is possible
that the higher order transitions are biased to the excited gas close
to a central starburst and do not necessarily trace the entire
molecular gas reservoir seen in \aco.  Also, the conversion factor
('$\alpha$') to derive molecular (H$_2$) masses from measured CO
luminosities has mostly been estimated for the \aco\ line (e.g.,
Downes \& Solomon 1998; Weiss et al.\ 2001).  Observing \aco\ has the
additional advantage that properties of the highly redshifted sources
can be {\em directly} compared to the molecular gas properties of
nearby (starburst) galaxies which are predominantly mapped in the
\aco\ transition.

These are main motivations for observing the \aco\ ground--state
transition.  However, due to the faintness of the line and the
bandwidth limitations of current radio telescopes, such high--$z$
\aco\ observations only exist for two QSOs and two radio galaxies to
date.  All of these observations have been obtained with radio
interferometers operating at centimeter wavelengths -- the NRAO Very
Large Array (VLA; Papadopoulos et al.\ 2001; Carilli et al.\ 2002a;
Greve et al.\ 2004) and the Australia Telescope Compact Array (ATCA;
Klamer et al.\ 2005). In particular, due to the bandwidth limitations
of the VLA, obtaining better constraints on the spectral line shape
and total flux of the \aco\ transition is desirable even for already
detected sources. Today's largest single dish telescopes, such as the
NRAO Green Bank Telescope (GBT)\footnote{The Green Bank Telescope is a
  facility of the National Radio Astronomy Observatory, operated by
  Associated Universities, Inc., under a cooperative agreement with
  the National Science Foundation.}  and the MPIfR Effelsberg
telescope\footnote{The Effelsberg telescope is a facility of the
  Max-Planck-Gesellschaft (MPG), operated by the Max-Planck-Institut
  f{\"u}r Radioastronomie (MPIfR).}, can eliminate some of those
issues due to their larger spectral bandwidths.

Here we report on first observations of \aco\ in three CO--bright
high--$z$ QSOs, which also exhibit ultra--luminous IR emission, using
the GBT and the Effelsberg telescope. In Section 2, we describe our
observations. Section 3 summarizes our results on the individual
objects (BR\,1202-0725 at $z=4.69$, PSS\,J2322+1944 at $z=4.12$, and
APM\,08279+5255 at $z=3.91$).  Section 4 provides an analysis and
discussion, and Section 5 closes with a summary of our results.  We
assume a standard $\Lambda$ cosmology throughout, with $H_0 =
71\,$\kms\ Mpc$^{-1}$, $\Omega_M = 0.27$, and $\Omega_{\Lambda} =
0.73$ (Spergel et al.\ 2003).

\section{Observations}

\subsection{GBT} \label{gbtobs}

Observations of the three targets were carried out with the GBT during
12 observing runs between October 2004 and April 2005 with a total
observing time of 89.5\,h.  The total/on--source observing times were
31.5/20\,h for BR\,1202-0725, 23/15\,h for PSS\,J2322+1944, and
35/22\,h for APM\,08279+5255.  During all runs, 3C147 and 3C286 were
used as primary/flux calibrators. For spectral line calibration, we
observed IRC+10216, DR21, Orion\,IRc2, Orion\,HC, and W3OH\,H$_2$O. As
secondary/pointing calibrators, 3C273, J1256-0547 (for BR\,1202-0725),
J2253+1608 (for PSS\,J2322+1944), J0753+538, and J0824+5552 (for
APM\,08279+5255) were utilized. The pointing accuracy, determined by
continuum cross scans of nearby sources, was typically $\sim$3\arcsec\
(reaching $\lesssim$1\arcsec\ under the best conditions).  We estimate
the calibration to be accurate to 10--15\%.  As the \aco\ transition
at 115.2712\,GHz is redshifted into the $K$ band for all three targets
(BR\,1202-0725: $\nu_{\rm obs} = 20.2450\,$GHz, PSS\,J2322+1944:
$\nu_{\rm obs} = 22.5258\,$GHz, APM\,08279+5255: $\nu_{\rm obs} =
23.4663\,$GHz), the dual--beam, dual--polarization 18--26\,GHz
receiver was used for all observations. The beam size of the GBT at
our observing frequencies is $\sim$32--36$''$ ($\sim$225--250\,kpc at
$z\simeq 4$), i.e., much larger than our targets.  The two beams have
a fixed separation of 178.8$''$ in azimuth direction. Two different
spectrometer setups were used; half of the observing runs were
executed in the first mode, and the other half of the runs in the
second mode. The first mode features two simultaneous intermediate
frequencies (IFs) with a bandwidth of 800\,MHz
($\sim$10,000--12,000\,\kms\ at our observing frequencies) and 2048
channels each, resulting in a spectral resolution of 391\,kHz
($\sim$5\,\kms). The \aco\ line was always centered in the first
800\,MHz IF. The second mode has one IF with 200\,MHz
($\sim$2,500--3,000\,\kms) and 16384 channels, resulting in a spectral
resolution of 12\,kHz ($\sim$0.15\,\kms).  BR\,1202-0725 and
APM\,08279+5255 were observed with both setups, while PSS\,J2322+1944
was only observed during runs that used the second setup.  The spectra
taken in each mode were examined separately and then combined.  The
ON--OFF position switching mode was used, i.e., the target was
observed alternatively with the two telescope beams, and the
off--source beam was always monitoring the sky background in parallel.
The beam switching frequency was once every 60--120 seconds depending
on the observing run. The weather was excellent for the winter nights
with typical zenith system temperatures of $T_{\rm sys} = 22-35$\,K on
a $T_{\rm A}^{\star}$ scale.  On April 19 and 20, 2005, the system
temperatures were significantly higher (typically $T_{\rm sys} =
50-60$\,K).

\begin{figure}
\epsscale{0.8}
\plotone{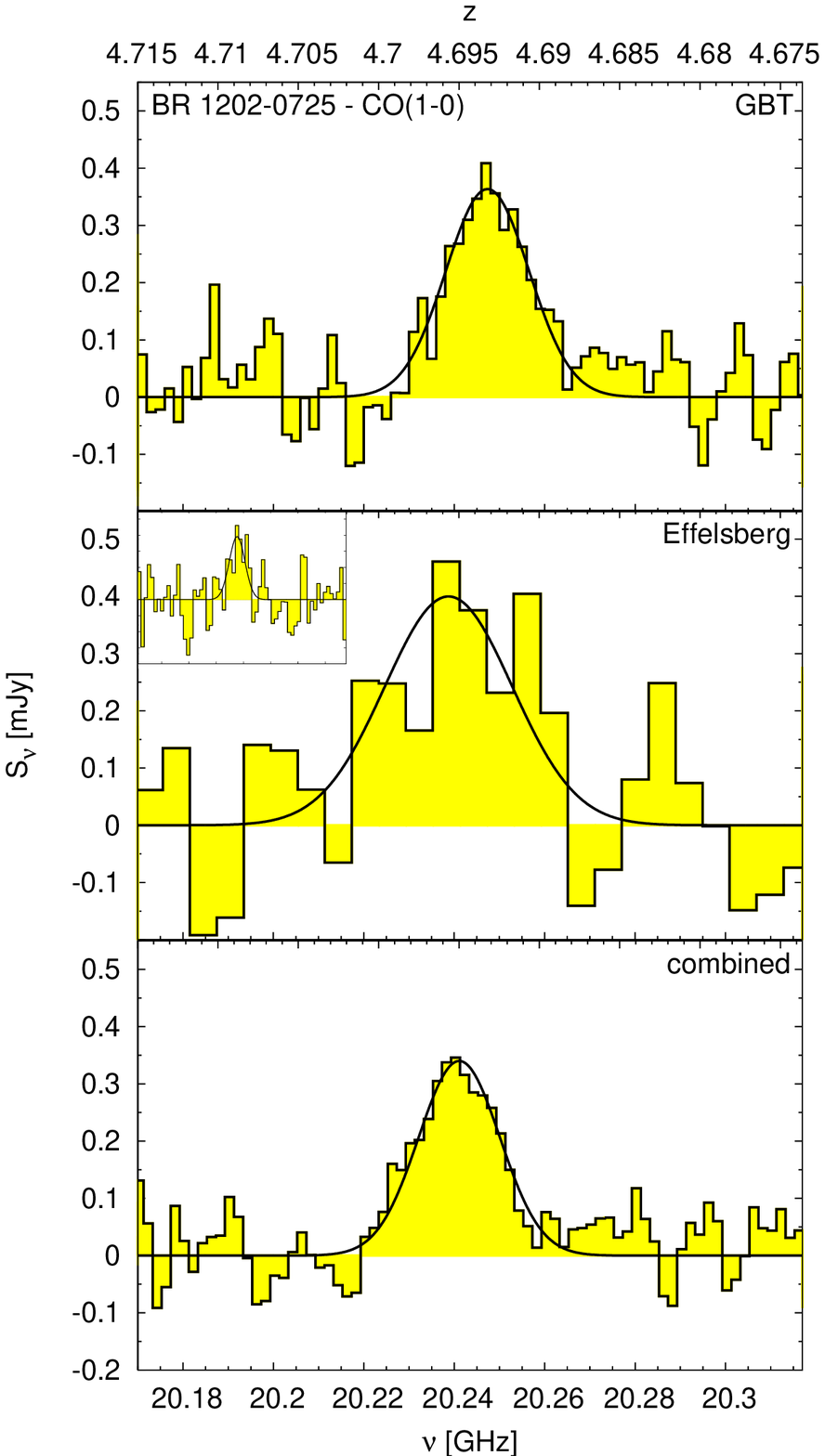}
\caption{GBT ({\em top}) and Effelsberg ({\em middle}) spectra of \aco\ 
emission from BR\,1202-0725. 
The {\em bottom} panel shows the combined spectrum. The GBT spectrum
has been smoothed to a resolution of 2.00\,MHz (30\,\kms). The rms per
channel is $\sim$75\,$\mu$Jy. The Effelsberg spectrum has been
smoothed to a resolution of 5.97\,MHz (88\,\kms). The rms per channel
is $\sim$155\,$\mu$Jy. For illustration, the inset shows the full
spectral range of the Effelsberg spectrum that was used for spectral
baseline fitting (width: $\sim$380\,MHz). The combined spectrum has
been smoothed to a resolution of 2.00\,MHz (30\,\kms). The rms per
channel is $\sim$70\,$\mu$Jy. The thin black lines show Gaussian fits
to the data (see Tab.\ \ref{tab-03}).
\label{fig1}}
\end{figure}

\begin{figure}
\epsscale{0.8}
\plotone{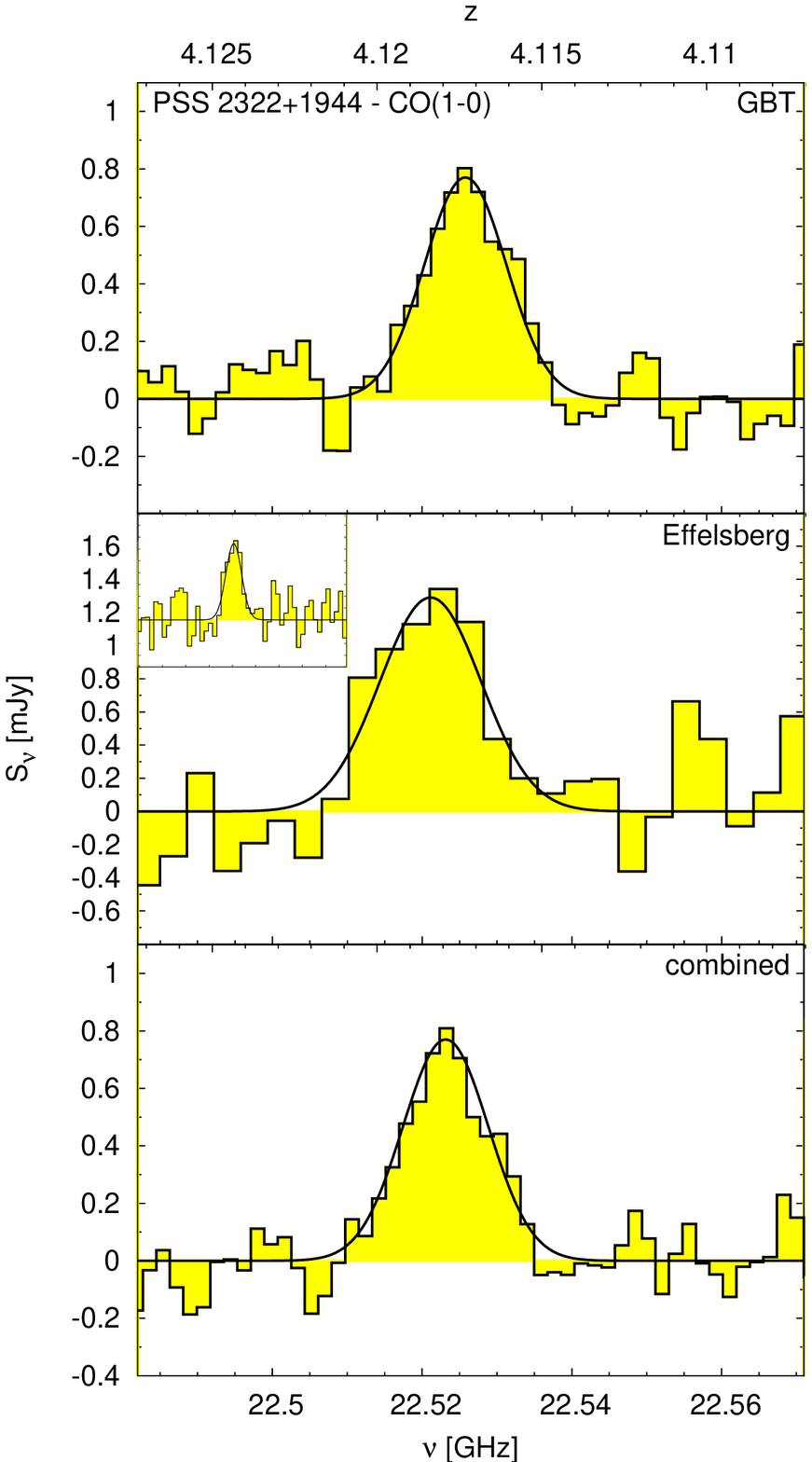}
\caption{GBT ({\em top}) and Effelsberg ({\em middle}) spectra of \aco\ 
emission from PSS\,J2322+1944.
The {\em bottom} panel shows the combined spectrum. The GBT spectrum
has been smoothed to a resolution of 1.80\,MHz (24\,\kms). The rms per
channel is $\sim$140\,$\mu$Jy. The Effelsberg spectrum has been
smoothed to a resolution of 3.60\,MHz (48\,\kms). The rms per channel
is $\sim$380\,$\mu$Jy.  For illustration, the inset shows the full
spectral range of the Effelsberg spectrum that was used for spectral
baseline fitting (width: $\sim$180\,MHz). The combined spectrum has
been smoothed to a resolution of 1.80\,MHz (24\,\kms). The rms per
channel is $\sim$125\,$\mu$Jy. The thin black lines show Gaussian fits
to the data (see Tab.\ \ref{tab-03}).
\label{fig2}}
\end{figure}

\begin{figure}
\epsscale{0.8}
\plotone{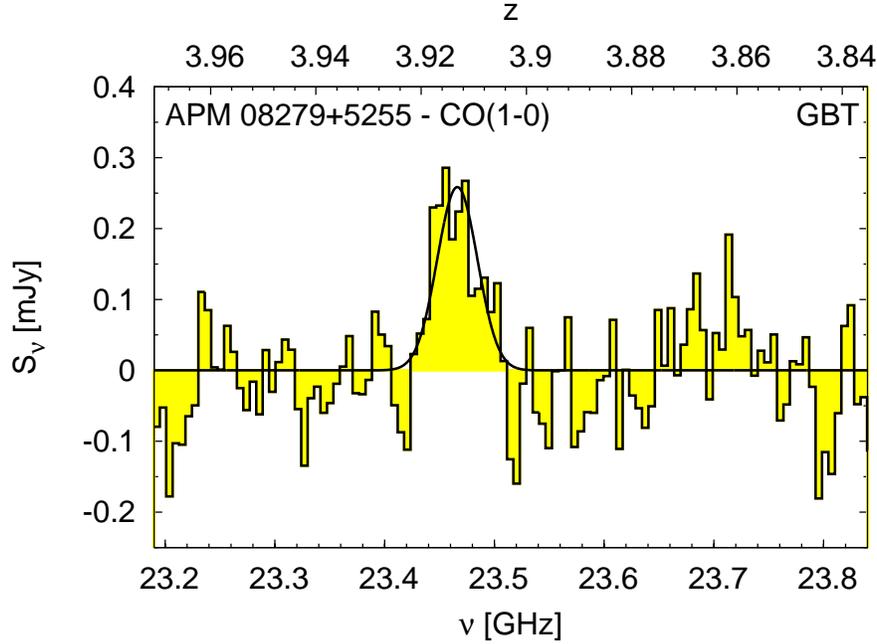}
\caption{GBT spectrum of \aco\ emission from APM\,08279+5255. 
The spectrum has been smoothed to a resolution of 5.86\,MHz
(75\,\kms). The rms per channel is $\sim$65\,$\mu$Jy. The thin black
line shows a Gaussian fit to the data (see Tab.\ \ref{tab-03}).
\label{fig3}}
\end{figure}

For data reduction, the aips++\footnote{aips2.nrao.edu} package and
the new GBT IDL software\footnote{gbtidl.sourceforge.net} were used,
providing consistent results.  The spectra of BR\,1202-0725,
PSS\,J2322+1944 and APM\,08279+5255 were binned to 2.00\,MHz
(30\,\kms), 1.80\,MHz (24\,\kms), and 5.86\,MHz (75\,\kms), reaching
rms noise values of $\sim$75\,$\mu$Jy, $\sim$140\,$\mu$Jy, and
$\sim$65\,$\mu$Jy.  A linear baseline was subtracted from the
PSS\,J2322+1944 spectrum to remove continuum fluxes and
atmospheric/instrumental effects. For BR\,1202-0725 and
APM\,08279+5255, polynomials of order 2 have been used to remove a
very wide ($\gg$100\,MHz, i.e., much broader than the width of the CO
line) `bending` of the baselines.  Only the frequency range that was
used to define the spectral baselines is used for the final GBT
spectra.  APM\,08279+5255 was observed using both the 200\,MHz and
800\,MHz bandwidth setups.  While the \aco\ line was clearly detected
with both setups, the spectrum obtained with the narrow--band
high--resolution 200\,MHz setup shows significant baseline problems,
as the continuum level is significantly different on the red and blue
sides of the line.  Therefore, we only use the results obtained with
the wide 800\,MHz bandwidth setup (corresponding to 15\,h on--source,
and covering a velocity range of $\sim$8,000\,\kms) for the final
spectrum.  In addition to simple ON-OFF combinations and subtraction
of low--order polynomial spectral baselines, we also followed the
scheme proposed by Vanden Bout et al.\ (2004) for reconstructing the
temporal baseline variations, however found that this did not improve
our final results.  The reduced spectra were read into
GILDAS/CLASS\footnote{www.iram.fr/IRAMFR/GILDAS} to write out the
final combined spectrum tables. The final spectra are shown in the top
panels of Figs.\ \ref{fig1} and \ref{fig2} and in Fig.\ \ref{fig3}. We
note that the spectrum of BR\,1202-0725 is limited to the central
$\sim$150\,MHz due to the fact that about half of the data were taken
with the narrower 200\,MHz setting.

\subsection{Effelsberg} \label{effobs}

Observations were carried out towards BR\,1202-0725 and
PSS\,J2322+1944 in January and February 2003 with a total observing
time of $\sim$40\,h (20\,h per source, corresponding to 8\,h
on--source each).  At 40\arcsec\ beam size, the pointing accuracy, as
determined by continuum cross scans of nearby sources, was better than
10\arcsec, with typical values of $\sim$5\arcsec. Calibrations for the
gain as well as the variation of the atmospheric opacity and zenith
distance were obtained from observations of 3C286 and NGC\,7027 (see
Ott et al. 1994 for reference fluxes), leading to a total formal
calibration uncertainty of $\sim$15\%.  We observed in beam switching
mode utilizing a rotating horn with a beam throw of 2\arcmin\ and a
switching frequency of $\sim$1\,Hz.  We have used a dual polarization
HEMT receiver for all observations.  The autocorrelator backend was
split into eight bands of 160\,MHz bandwidth and 128 channels each
that could individually be shifted in frequency by up to $\pm$250\,MHz
relative to the recessional velocity of the targets. The final spectra
cover a velocity range of $\sim$6,000\,\kms\ with channel spacings of
$\sim$16\,km\,s$^{-1}$.  We achieved single channel system
temperatures of 65\,K and 85\,K on a $T_{\rm A}^{\star}$ scale.  After
combination of both orthogonal linear polarizations, this leads to a
$T_{\rm sys}$ of 47\,K and 60\,K.

The GILDAS/CLASS package was used for data reduction.  All spectral
baselines are of good quality and only first order polynomial
baselines had to be subtracted. The final spectra of BR\,1202-0725 and
PSS\,J2322+1944, are shown in the middle panels of Figs.\ \ref{fig1}
and \ref{fig2}.  These spectra were binned to 5.97\,MHz
(88\,km\,s$^{-1}$) and 3.60\,MHz (48\,km\,s$^{-1}$), reaching rms
noise values of $\sim$150\,$\mu$Jy and $\sim$380\,$\mu$Jy.  We note
that the latter do not show the full ranges used for spectral baseline
fitting. Those ranges ($\sim$380\,MHz for BR\,1202-0725, and
$\sim$180\,MHz for PSS\,J2322+1944) are illustrated by the insets in
the same figures.

\subsection{Combined spectra} \label{combined}

As BR\,1202-0725 and PSS\,J2322+1944 were observed with both
telescopes, we also created combined spectra of both results to
increase the signal--to--noise ratio and average out part of the
calibrational uncertainties. The GILDAS/CLASS package was used to
reprocess and combine the final spectra of both telescopes. The final
unbinned GBT and Effelsberg spectra were regridded to a common
velocity resolution and baseline--subtracted before combination.  In
the combination, spectra were weighted with their respective rms.  The
combined spectra of BR\,1202-0725 and PSS\,J2322+1944, as shown in the
bottom panels of Figs.\ \ref{fig1} and \ref{fig2}, were binned to a
resolution of 2.00\,MHz (30\,km\,s$^{-1}$) and 1.80\,MHz
(24\,km\,s$^{-1}$), reaching rms noise values of $\sim$70\,$\mu$Jy and
$\sim$125\,$\mu$Jy.

\section{Results}

The sources in our study are the three CO--brightest high--redshift
QSOs that can currently be observed in the \aco\ transition.  We
obtained detections for all our targets.  CO line luminosities
$L'_{\rm CO}$ (in K \kms pc$^2$) were derived utilizing:
\begin{equation}
  L'_{\rm CO} = 3.25 \times 10^7 \times I \times \nu_{\rm obs}^{-2} \times D_{\rm L}^2 \times (1+z)^{-3}
\end{equation}

where $I$ is the velocity--integrated \aco\ line flux in Jy \kms,
$D_{\rm L}$ is the luminosity distance in Mpc, and $\nu_{\rm obs}$ is
the observed frequency in GHz (Solomon et al.\ 1992).  For the systems
with known lensing magnification factor $\mu_{\rm L}^{\rm CO}$,
$L'_{\rm CO}$ has to be divided by that factor in order to get the
intrinsic CO luminosity of the discussed target.  A conversion factor
$\alpha = 0.8$\,M$_{\odot}\,$/K \kms pc$^2$ to convert $L'_{\rm
CO(1-0)}$ to $M_{\rm gas}({\rm H_2})$ is assumed throughout, as
applicable for local ultra--luminous infrared galaxies
(ULIRGs)/starbursts (Downes \& Solomon 1998).  All observational
results are summarized in Table \ref{tab-03}.  Derived CO luminosities
and gas masses as well as FIR luminosities from the literature are
given in Table \ref{tab-04}.

\subsection{BR\,1202-0725} \label{br1202}

\subsubsection{Previous results}

BR\,1202-0725 ($z = 4.69$) was detected in multiple CO transitions
before (see Table \ref{tab-01}), but not in \aco.  This optically
bright, radio--quiet QSO has the curious property that the optical QSO
is a single source, but the mm continuum and CO line observations show
a double source with a separation of about 4$''$ (Omont et al.\ 1996;
Guilloteau et al.\ 1999; Carilli et al.\ 2002b).  This double
morphology may indicate a pair of interacting objects separated by
only 28\,kpc (Yun et al.\ 2000; Carilli et al.\ 2002b).  An alternate
explanation would be a double starburst system composed of a QSO
(southern source) and a dust--obscured, Ly--$\alpha$ - emitting
companion (northern source, Hu et al.\ 1996), which is ionized by the
strong QSO.  Recently, Klamer et al.\ (2004) suggested that the nature
of the double source might be due to jet--induced star--formation,
where the northern component corresponds to a radio hot spot.  It has
been discussed that the quasar activity in this system may be
triggered by gravitational interaction with the companion.  Using
typical conversion factors for a ULIRG (Downes \& Solomon 1998), the
total molecular gas mass derived from the CO luminosity exceeds the
dynamical mass of the system (Carilli et al.\ 2002b), and as a remedy
for this inconsistence, gravitational lensing has been suggested.
However, the same authors argue that masses and velocity widths of the
components are very different, therefore it appears unlikely that the
multiple components are different lensed images of one source.

\subsubsection{New observations}

The final \aco\ spectra of BR\,1202-0725 are shown in Fig.\ \ref{fig1}
({\em top panel}: GBT, {\em middle panel}: Effelsberg, {\em bottom
panel}: combined spectrum).  Gaussian fitting to the line profile of
the combined spectrum results in a peak flux density of 0.34 $\pm$
0.03\,mJy and a FWHM of 333 $\pm$ 30\,\kms. The integrated \aco\ line
flux is 0.120 $\pm$ 0.010\,Jy \kms. This agrees well with the
extrapolated value of 0.123 $\pm$ 0.013\,Jy \kms\ derived from the
\bco\ flux (Carilli et al.\ 2002b) under assumption of fully
thermalized and optically thick CO emission (see also discussion in
Sect.\ 4.1). The width of the \aco\ line is consistent within the
error bars with an average of the higher--$J$ transitions in the
literature ($\sim 290\,$\kms, see Table
\ref{tab-01}). The derived redshift of 4.6949 $\pm$ 0.0003
is in good agreement with the \eco\ redshift (Omont et al.\ 1996,
northern and southern components combined). The structure of this
source is unresolved at the resolution of our measurements; all line
parameters are in good agreement with the higher--$J$ lines adding up
both components.

The derived CO line luminosity of $1.0 \times 10^{11}\,$K \kms pc$^2$
results in an H$_2$ gas mass of $8.1 \times 10^{10}$\,M$_{\odot}$ (see
also Tab.\ \ref{tab-04}).

\subsection{PSS\,J2322+1944} \label{pss2322}

\subsubsection{Previous results}

The $z=4.12$ QSO PSS\,J2322+1944 is an IR--luminous high--redshift
source (see Table \ref{tab-04}) that is known to exhibit strong CO
line emission in multiple transitions (see Table \ref{tab-01}).
Optical imaging and spectroscopy reveal a double source structure, and
the two components are separated by about $1.5''$.  The spectra of
both peaks are essentially identical, consistent with strong
gravitational lensing (optical magnification factor $\mu_{\rm L}^{\rm
  opt} = 3.5$) by an intervening foreground galaxy (Carilli et al.\
2003). In high--resolution VLA images, Carilli et al.\ (2003) find a
molecular Einstein Ring with a diameter of $1.5''$ in \bco\ line
emission, which can be modeled as a circumnuclear star--forming disk
with a radius of 2.2\,kpc (CO magnification factor $\mu_{\rm L}^{\rm
  CO} \sim 2.5$).  The derived intrinsic star--formation rate (SFR) is
of order 900\,M$_{\odot}$\,yr$^{-1}$. PSS\,J2322+1944 exhibits strong,
non--thermal (synchrotron) radio continuum emission at 1.4\,GHz, and
the rest--frame radio--to--IR spectral energy distribution (SED)
resembles that of local nuclear starburst galaxies like M82 (Cox et
al.\ 2002). This QSO is the fourth high--$z$ CO emitter to be detected
in [C\ts {\scriptsize I}] emission (Pety et al.\ 2004), providing
additional evidence for the presence of active star--formation in the
host galaxy.

\subsubsection{New observations}

The final \aco\ spectra of PSS\,J2322+1944 are shown in Fig.\
\ref{fig2} ({\em top panel}: GBT, {\em middle panel}: Effelsberg, {\em
bottom panel}: combined spectrum). From Gaussian fitting to the
combined spectrum, the peak line flux density is found to be 0.77
$\pm$ 0.07\,mJy, the line FWHM is 190 $\pm$ 14\,\kms, and the
integrated line flux is 0.155 $\pm$ 0.013\,Jy \kms. These values are
in good agreement with the \aco\ detection of Carilli et al.\ (2002a;
$S_\nu = 0.89 \pm 0.22\,$mJy, $\Delta V_{\rm FWHM} = 200 \pm
70\,$\kms, $I = 0.19 \pm 0.08\,$Jy \kms), although results for the
higher--$J$ CO transitions indicate a larger linewidth ($\Delta V_{\rm
FWHM} > 250\,$\kms, see Table \ref{tab-01}).  The Gaussian fit gives a
redshift of 4.1179 $\pm$ 0.0002.

The derived lensing--corrected CO line luminosity of $4.2 \times
10^{10}\,$K \kms pc$^2$ results in an H$_2$ gas mass of $3.4 \times
10^{10}$\,M$_{\odot}$ (see also Tab.\ \ref{tab-04}).

\subsection{APM\,08279+5255} \label{apm0827}

\subsubsection{Previous results}

APM\,08279+5255 is a strongly lensed, radio--quiet broad absorption
line (BAL) QSO at $z = 3.91$. Gravitational lens models of the QSO
continuum source suggest magnification by a factor of $\mu_{\rm
L}^{\rm opt} \sim 100$, and the image breaks up into three components
with a maximum separation of $0.4''$ (Ledoux et al.\ 1998; Ibata et
al.\ 1999; Egami et al.\ 2000). APM\,08279+5255 has been detected in
the mm and sub--mm dust continuum, revealing an apparent bolometric
luminosity of $\sim 5 \times 10^{15}\,$L$_\odot$ (Lewis et al.\
1998). A multi--transition CO study (see Table \ref{tab-01}) in
combination with detailed lens models appear to reveal a spatially
extended structure on a scale of at least 400\,pc (Lewis et al.\
2002), which is gravitationally magnified by a factor of $\mu_{\rm
L}^{\rm CO} = 7$.  The strength of the \ico\ emission indicates the
presence of hot dense gas with a kinetic temperature of approximately
200\,K (Downes et al.\ 1999).

From VLA imaging of the CO ground--state transition at a linear
resolution of $2.25''$, Papadopoulos et al.\ (2001) report the
detection of an extended, low--excitation molecular gas reservoir
around the compact nucleus which extends over a scale of $7''$
($\sim$30\,kpc). The integrated brightness temperature of this
extended domain appears to be of the same order of magnitude as that
of the nuclear region. This extended reservoir would be well within
our 32$''$ GBT beam.  Papadopoulos et al.\ (2001) do not derive the
total flux in the extended reservoir.  For the central $\sim$1$''$
(corresponding to $\sim$7.2\,kpc at the source redshift), which they
call "the nuclear \aco\ emission", they find an integrated flux of
0.150 $\pm$ 0.045\,Jy \kms. The VLA bandpass used by Papadopoulos et
al.\ (2001) has an effective bandwidth of $\sim$45\,MHz, or
$\sim$575\kms\ at the \aco\ line frequency.  Therefore, assuming that
the peak in their Figure 1c is 6.5$\sigma = 260\,\mu$Jy beam$^{-1}$
($\sigma=40\,\mu$Jy beam$^{-1}$), we consistently derive an integrated
flux over the peak of 0.15\,Jy \kms\ beam$^{-1}$. However,
Papadopoulos et al.\ (2001) suggest that this peak of 'nuclear
emission' sits on a broad, extended plateau (see their Figure 1c). If
we assume that the extended reservoir component is traced by their
2$\sigma$ ($80\,\mu$Jy beam$^{-1}$) contour, it has a width of
$\sim$8$''$$\times$3$''$, or $\sim$4.7 beam areas at the given
resolution of 2.25$''$$\times$2.25$''$. This corresponds to a flux of
0.22\,Jy \kms\ for the extended emission. The same estimate for their
3$\sigma$ ($120\,\mu$Jy beam$^{-1}$) contour gives a width of
$\sim$7$''$$\times$2.25$''$, or $\sim$3.1 beam areas. This again
corresponds to a flux of 0.22\,Jy \kms\ (note that both these
estimates are lower limits for the extended emission). The total flux
in the \aco\ map of Papadopoulos et al.\ (2001) is therefore estimated
to be $\gtrsim$0.37\,Jy \kms. The same authors also suggest that this
reservoir breaks up into multiple components at higher resolution; in
their \bco\ observations at 0.5$''$ resolution, they claim to find two
emitting regions 2--3$''$ distant from the central region -- if real,
these could be companion galaxies which are not individually resolved
in the extended \aco\ reservoir.

\subsubsection{New observations}

The final \aco\ spectrum of APM\,08279+5255 is shown in Fig.\
\ref{fig3}.  We derive a \aco\ peak flux of 0.26 $\pm$ 0.04\,mJy and
an FWHM of 556 $\pm$ 55\,\kms\ from our Gaussian fit. This results in
an integrated line flux of 0.152 $\pm$ 0.020\,Jy \kms, which is in
good agreement with the value of 0.150 $\pm$ 0.045\,Jy \kms\ found by
Papadopoulos et al.\ (2001) for the central $\sim$1$''$ ("the nuclear
\aco\ emission"). However, our result is inconsistent with the
integrated flux of 0.37\,Jy \kms\ given by our estimate of their full
\aco\ reservoir (see previous subsection). Such a high flux is clearly
ruled out by our GBT observations. We thus find no evidence for a
luminous extended halo, which would be well within our 32$''$ beam,
and cannot confirm the existence of bright companion galaxies.  The
derived FWHM velocity width of our \aco\ detection is in agreement
with single--dish observations of the \fco, \jco, and \kco\
transitions obtained with the IRAM 30\,m telescope ($\sim$500\,\kms,
Weiss et al.\ 2006, in prep.) and IRAM Plateau de Bure interferometer
observations of the \dco\ transition (480 $\pm$ 35\,\kms, Downes et
al.\ 1999, see Table \ref{tab-01}).  Our derived \aco\ redshift of
3.9122 $\pm$ 0.0007 is in good agreement with previous results (3.9114
$\pm$ 0.0003 for the 4$\to$3 transition). The derived
lensing--corrected CO line luminosity of $1.4 \times 10^{10}\,$K \kms
pc$^2$ results in an H$_2$ gas mass of $1.1 \times
10^{10}$\,M$_{\odot}$ (see also Tab.\ \ref{tab-04}).

\section{Analysis and Discussion}

\subsection{LVG modeling} \label{LVG}

\begin{figure}
\epsscale{0.7}
\plotone{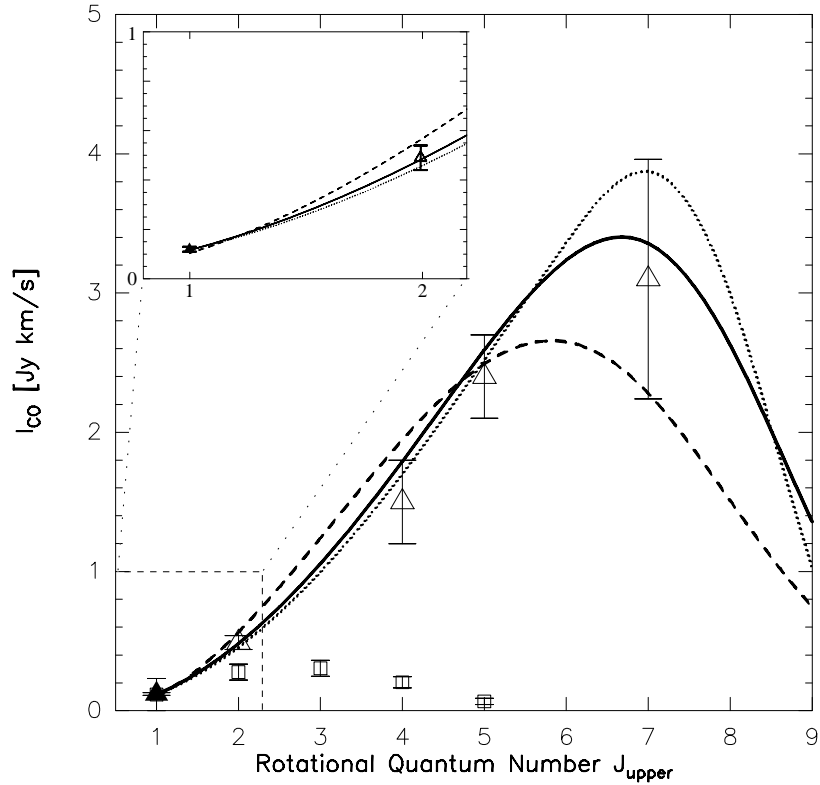}
\caption{
CO ladder and LVG models (based on all $J > 1$ transitions) for
BR\,1202-0725. The inset shows a zoomed--in version of the $J = 2$, $J
= 1$ region (as indicated by the dashed box). The filled triangle is
the \aco\ data point based on the combined spectrum. Data for the
higher--$J$ CO transitions (open triangles) are taken from the
literature (see Table \ref{tab-01}). For comparison, we also show data
for the inner disk of the Milky Way (open squares, Fixsen et al.\
1999), normalized to the \aco\ flux of BR\,1202-0725. The kinetic
temperature $T_{\rm kin}$ and gas density $\rho_{\rm gas}({\rm H_2})$
are treated as free parameters in this study. Three representative
models are shown: Model 1 (solid line) assumes $T_{\rm kin} = 60\,$K
and $\rho_{\rm gas}({\rm H_2}) = 10^{4.1}\,$cm$^{-3}$, and gives the
overall best fit to all transitions. Model 2 (dashed line) assumes
$T_{\rm kin} = 120\,$K and $\rho_{\rm gas}({\rm H_2}) =
10^{3.7}\,$cm$^{-3}$, while model 3 (dotted line) assumes $T_{\rm kin}
= 30\,$K and $\rho_{\rm gas}({\rm H_2}) = 10^{4.6}\,$cm$^{-3}$.
\label{fig6}}
\end{figure}

To investigate how much of the \aco\ emission in our target QSOs is
associated with the molecular gas reservoirs detected in the
higher--$J$ CO transitions, we have used spherical, one--component
large velocity gradient models (LVG).  All these LVG calculations use
the collision rates from Flower \& Pineau des Forets (2001) with an
ortho/para H$_2$ ratio of 3 and a CO abundance per velocity gradient
of [CO]/$\nabla v = 1 \times 10^{-5}\,{\rm pc}\,$(\kms)$^{-1}$ (e.g.\
Weiss et al.\ 2005). Models were fitted to those lines {\em above} the
\aco\ transition listed in Table \ref{tab-01} for each source. The
turnover of the CO line SED (and therefore the slope beyond the
turnover) is not well determined. Thus, a large degeneracy exists
between the kinetic gas temperature $T_{\rm kin}$ and density
$\rho_{\rm gas}({\rm H_2})$, the two main free parameters in our
study.  As an example, Figure \ref{fig6} shows data of all transitions
and three representative models for BR\,1202-0725: Model 1 (solid
line) assumes $T_{\rm kin} = 60\,$K and $\rho_{\rm gas}({\rm H_2}) =
10^{4.1}\,$cm$^{-3}$, and gives the overall best fit to all
transitions. Models 2 (dashed line) and 3 (dotted line) are shown as a
representation of the parameter space allowed by the data within the
error bars. Model 2 assumes $T_{\rm kin} = 120\,$K and $\rho_{\rm
gas}({\rm H_2}) = 10^{3.7}\,$cm$^{-3}$, while model 3 assumes $T_{\rm
kin} = 30\,$K and $\rho_{\rm gas}({\rm H_2}) = 10^{4.6}\,$cm$^{-3}$.
The LVG predicted \aco\ flux of the different models based on the
$J>1$ CO transitions is fairly well constrained by the solutions.
Most of our calculated models suggest that the CO emission is close to
thermalized up to the \dco\ transition and optically thick ($\tau
\gtrsim 5$), which implies that the LVG predicted \aco\ line fluxes
are similar to those we would derive by assuming a $\nu^2$ scaling of
the line flux densities from the mid--$J$ CO transitions.  The
predicted LVG integrated flux ranges of the \aco\ transition are
0.10--0.12\,Jy \kms\ (BR\,1202-0725), 0.20--0.23\,Jy \kms\
(PSS\,J2322+1944), and 0.13--0.20\,Jy \kms\ (APM\,08279+5255) and are
in good agreement with our observations for all sources.  As described
in Sect.\ 3.2.2, the observed \aco\ linewidth in our new observations
and in the VLA spectrum (Carilli et al.\ 2002a) of PSS\,J2322+1944 is
lower than that of the higher--$J$ transitions, thus the integrated
model fluxes are a bit higher than our result. If we compare the peak
fluxes only, we obtain a good agreement.

Our observations and models are in agreement with the assertion that
{\em all} observed \aco\ flux density is associated with the highly
excited molecular gas seen in the high--$J$ CO lines. We thus find no
evidence for an additional luminous, more extended low surface
brightness gas component surrounding the central region of our target
QSOs, in contrast to what has been suggested previously for
APM\,08279+5255 (Papadopoulos et al.\ 2001).  Given the accuracy of
our measurements, we conclude that at most 20--30\% of the \aco\
luminosity may be associated with such a diffuse component.  We note
however that, if the $L'_{\rm CO}$ to $M_{\rm gas}({\rm H_2})$
conversion factor ('$\alpha$') for a faint extended component were
higher (e.g.\ Galactic), a higher H$_2$ mass may be hidden in such an
extended component.

\subsection{Correlations of high and low redshift galaxies} 
\label{gaosolomon}

CO results for our three high--redshift IR--luminous QSOs are
summarized in Tab.~\ref{tab-03} and \ref{tab-04}.  Our sample consists
of all three high--$z$ QSOs for which the CO ground--state transition
has been detected to date, and covers $\sim$20\% of the CO--detected
$z>2$ quasars.  As discussed above, the \aco\ transition provides the
best information of the total amount of molecular gas in a system,
quantified by $L'_{\rm CO(1-0)}$. For most QSOs/galaxies detected in
CO at high $z$, \cco\ and \dco\ are the lowest transitions that have
been detected to date. Based on our results in the previous paragraph,
however, we can now estimate their $L'_{\rm CO(1-0)}$ by assuming
constant brightness temperature for all transitions from \aco\ to
\dco\ (i.e., $L'_{\rm CO}$ is the same for those transitions).  This
assumption is also in agreement with observations towards the radio
galaxy 4C60.07 (Greve et al.\ 2004). However, it is important to note
that all high--$z$ sources were selected via higher--$J$ CO
transitions, which could, in principle, introduce a bias towards
highly excitated starburst environments.

Following Sanders et al.\ (1991), Gao \& Solomon (2004) found a
non--linear relation between the logarithms of the FIR luminosity
$L_{\rm FIR}$ and the \aco\ line luminosity $L'_{\rm CO}$ for a sample
of local spiral galaxies, luminous, and ultra--luminous infrared
galaxies. Their sample consists mostly of galaxies whose FIR
luminosity is powered by star formation only.  Our \aco\ observations
and LVG models suggest that $L'_{\rm CO}$ can be estimated for all
CO--detected high--$z$ sources with some degree of confidence, even if
only observations of higher--$J$ CO transitions exist for most of
these sources. To put this result into context, we now aim to discuss
the $L'_{\rm CO}$--$L_{\rm FIR}$ relation and its implications for
some selected samples of galaxies at low and high redshift.

\begin{figure}
\epsscale{1.0}
\plotone{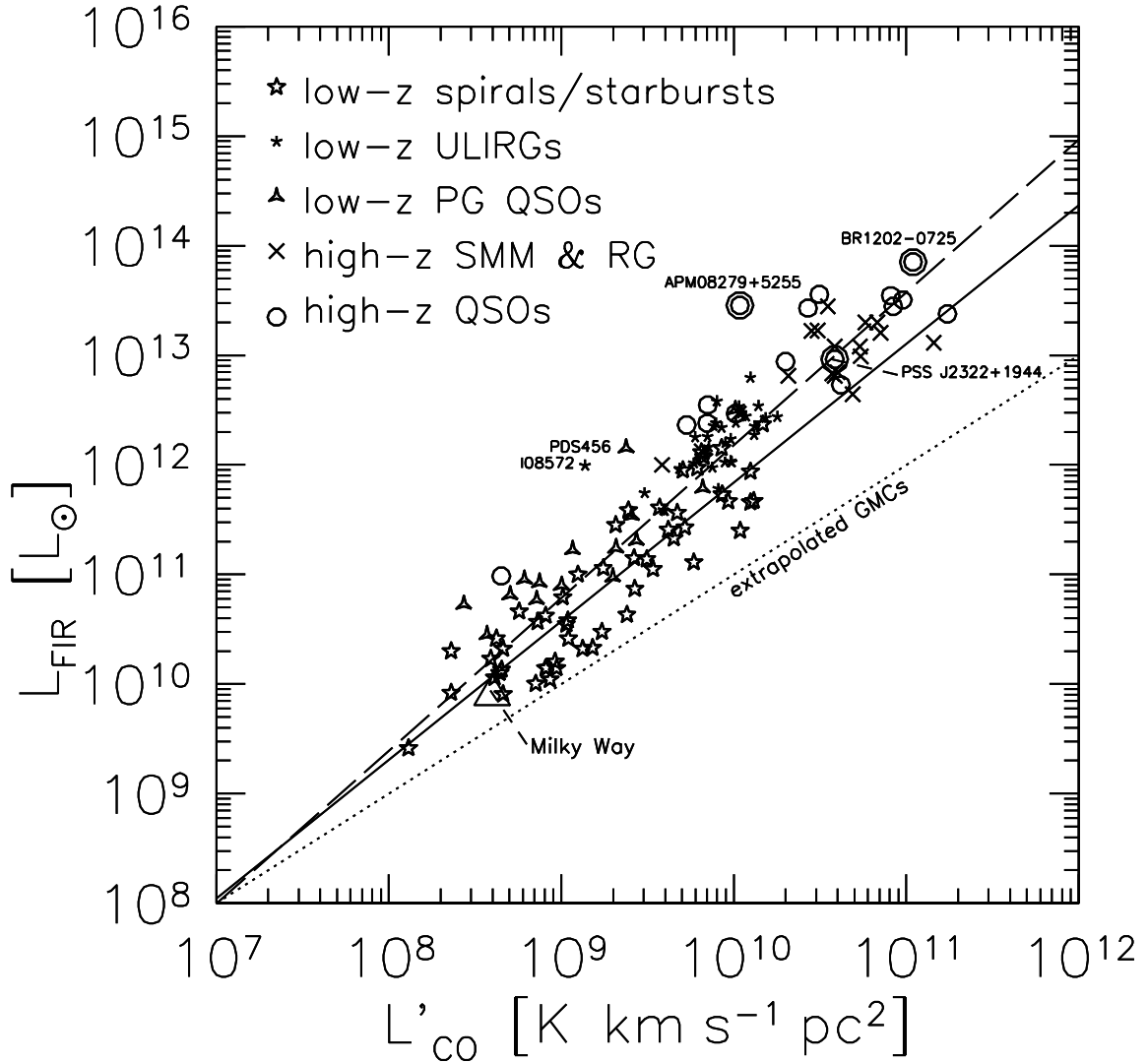}
\caption{Comparison of velocity--integrated CO line luminosity with
  FIR luminosity for a sample of low--$z$ spiral and starburst
  galaxies from Gao \& Solomon (2004), the ULIRGs from Solomon et al.\
  (1997), $z < 0.2$ Palomar--Green (PG) QSOs from Alloin et al.\
  (1992), Evans et al.\ (2001), and Scoville et al.\ (2003),
  extrapolated Galactic molecular clouds (GMCs) from Mooney \& Solomon
  (1988), the Milky Way (Fixsen et al.\ 1999), and for high--$z$
  sub--millimeter galaxies (SMM), radio galaxies, and QSOs from the
  literature (see Solomon \& Vanden Bout 2005) with respect to our new
  results.  All FIR luminosities are re--derived as described in
  Carilli et al.\ (2005). All data are corrected for gravitational
  lensing. The solid line is a straight--line least squares fit to the
  Gao \& Solomon (2004) sample, corresponding to log$(L_{\rm FIR}) =
  (1.26 \pm 0.08) \times {\rm log} (L_{\rm CO})-0.81$.  The dotted
  line shows the fit to the GMCs from Mooney \& Solomon (1988),
  corresponding to a power law index of $0.96 \pm 0.08$.  The dashed
  line is a fit to all high--$z$ sources, the Gao \& Solomon (2004)
  data, the Solomon et al.\ (1997) ULIRGs, and the PG QSOs,
  corresponding to log$(L_{\rm FIR}) = (1.39 \pm 0.05) \times {\rm
    log} (L_{\rm CO})-1.76$.\label{fig7}}
\end{figure}

Figure \ref{fig7} shows the relationship between log($L_{\rm FIR}$)
and log($L'_{\rm CO}$) for our three targets, all other high--$z$ CO
detections (except TN\,J0924-2201, which does not have a measured FIR
luminosity), the $z < 0.2$ Palomar--Green (PG) QSOs from Alloin et
al.\ (1992), Evans et al.\ (2001), and Scoville et al.\ (2003)
including PDS\,456 (Yun et al.\ 2004), local ($z < 0.3$) ULIRGs from
Solomon et al.\ (1997), and the local galaxy sample ($z < 0.1$) of Gao
\& Solomon (2004). All FIR luminosities for the high--$z$ sources are
re--derived as described by Carilli et al.\ (2005) unless stated
otherwise. Lensing magnification factors were taken into account. In
addition, data for the Milky Way disk (Fixsen et al.\ 1999) and an
(extrapolated) relation for Galactic molecular clouds (Mooney \&
Solomon 1988) are given for comparison. The solid line is a
straight--line least squares fit to the Gao \& Solomon (2004) sample,
corresponding to log$(L_{\rm FIR}) = (1.26 \pm 0.08) \times {\rm log}
(L_{\rm CO})-0.81$ (the power law index is 1.25 $\pm$ 0.08 in the
original publication).  Fitting the high--$z$ sources, the Gao \&
Solomon (2004) data, the Solomon et al.\ (1997) ULIRGs, and the PG
QSOs together, we find the relation log$(L_{\rm FIR}) = (1.39 \pm
0.05) \times {\rm log} (L_{\rm CO})-1.76$. Both relations are clearly
non--linear, as their power--law indexes are significantly larger than
unity (see also the corresponding discussion in Gao \& Solomon 2004).
We recover a Schmidt--Kennicutt law (power law index of 1.4, Kennicutt
1998a,b) for the larger, more heterogenous sample.  More importantly,
the high--$z$ sources seem to follow the same slope (within the
errors) as seen at low $z$, even though most high--$z$ sources harbor
central AGNs.  In addition, the high--$z$ QSOs and high--$z$
sub--millimeter and radio galaxies (likely lacking a luminous AGN)
statistically occupy the same area in the plot. The same behaviour is
found for the PG QSOs in comparison to the low--$z$
spirals/LIRGs/ULIRGs.  It is in general not clear that the FIR
emission comes mostly from star--formation, i.e.\ that it is not
strongly biased by active nuclei or the interstellar radiation field.
E.g., SED modeling of the $z=2.6$ quasar H1413+117 (the Cloverleaf)
reveals that only $\sim$20\% of its FIR luminosity is powered by star
formation (Solomon et al.\ 2003), but it is not an outlier in the
$L'_{\rm CO}$--$L_{\rm FIR}$ diagram.  In this context, the elevated
$L_{\rm FIR}$/$L'_{\rm CO}$ ratio in APM\,08279+5255 may be explained
by a combination of differential lensing and a relatively high
contribution of the AGN to $L_{\rm FIR}$ (see Rowan--Robinson 2000 for
models of the IR SED).  Outliers like APM\,08279+5255 are also found
at low $z$ (e.g.\ the QSO PDS\,456 and the warm ULIRG
IRAS\,08572+3915, see Fig.\ \ref{fig7}).  As the relative number of
such outliers is very small, this may also be explained by a brief
FIR--bright AGN phase (Yun et al.\ 2004).  However, it is likely that
the dominant energy source in most ULIRGs is an extreme starburst
rather than heating by a dust--enshrouded AGN (e.g., Solomon et al.\
1997; Downes \& Solomon 1998).  This picture is supported by the
finding that ULIRGs harbor large quantities of {\em dense} molecular
gas, which is more intimately involved with star formation than the
major fraction of the mostly diffuse CO (Gao \& Solomon 2004).  It has
to be kept in mind that the high--$z$ sources are highly selected, and
probably fulfill the Malmquist bias (i.e., the apparent increase in
the average $L'_{\rm CO}$ and $L_{\rm FIR}$ towards high redshift is
probably a consequence of the flux limitation in the sample, e.g.\
Sandage 1994). As an additional consequence of flux limitation in the
sample, lensing influences the results (i.e., stronger lensing
magnification allows to probe deeper, and therefore the intrinsic CO
luminosity drops).

We can also set an approximate lower limit on the duration of the
intense starburst phase. Considering the total H$_2$ masses and
star--formation rates in Table \ref{tab-04}, the depletion timescale
of the molecular gas is of order 10$^7$\,yr for all three QSOs under
the assumption of a constant SFR and 100\% star--forming efficiency.
This implies that the starburst itself can be relatively short--lived,
(and compact, as the dynamical time must be less than the starburst
lifetime) unless the molecular gas in which the star--formation occurs
can be re--supplied on timescales of $\sim 10^7$\,yr.

\section{Summary}

We have detected \aco\ emission in three QSOs at redshifts $3.9 < z <
4.7$ with the NRAO GBT and the MPIfR Effelsberg telescope.  From our
analysis of the resulting spectra of BR\,1202-0725, PSS\,2322+1944,
and APM\,08279+5255, we obtain the following results: \\

1. We derived lensing--corrected \aco\ line luminosities of $1.0
\times 10^{11}\,$K \kms pc$^2$ for BR\,1202-0725, $4.2 \times
10^{10}\,$K \kms pc$^2$ for PSS\,2322+1944, and $1.4 \times
10^{10}\,$K \kms pc$^2$ for APM\,08279+5255. These results are in good
agreement (within a factor of 2) with previous estimates of the total
CO luminosities based on the higher--$J$ CO transitions, consistently
providing very large $M_{\rm gas}({\rm H_2})$ of
$>10^{10}\,$M$_{\odot}$.

2. \aco\ fluxes predicted by one--component LVG models are in good
agreement with our observations.  Considering our modeling results,
the CO emission appears to be close to thermalized up to the 4$\to$3
transition in all cases. Thus, our observations show no indication of
a luminous extended, low surface brightness molecular gas component in
any of the high--redshift QSOs in our sample (cf.\ Papadopoulos et
al.\ 2001). In fact, all CO transitions are described very well by a
single gas component where all molecular gas is concentrated in a
compact circumnuclear region.  If such extended components were to
exist, our observations and models limit their contribution to the
overall luminosity to at most 20--30\%.

3. log($L_{\rm FIR}$) and log($L'_{\rm CO}$) appear to be correlated
for low--redshift galaxies over orders of magnitude in $L_{\rm FIR}$
including (U)LIRGs, and the significantly brighter sources found at
high--$z$ appear to follow the same general trend (see also Solomon \&
Vanden Bout 2005).  In particular, we find that the correlation shows
no significant difference between QSOs and systems without a luminous
AGN.

The observations presented herein demonstrate the feasibility of
detecting high--$z$ CO with 100\,m single--dish radio telescopes, and
highlight the physical implications of observing the ground--state
transition of this molecule towards massive galaxies at redshifts
greater than 4.

\acknowledgments

The National Radio Astronomy Observatory is operated by Associated
Universities Inc., under cooperative agreement with the National
Science Foundation. We would like to thank the staff at the GBT, in
particular J.\ Braatz, G.~I.\ Langston and R.~J.\ Maddalena, for their
extensive support and many helpful discussions. D.~R.\ acknowledges
support from the Deutsche Forschungsgemeinschaft (DFG) Priority
Programme 1177.  C.~C.\ acknowledges support from the
Max-Planck-Gesellschaft and the Alexander von Humboldt-Stiftung
through the Max-Planck-Forschungspreis.  We thank the referee for many
useful comments which helped to improve the manuscript.


\clearpage

\begin{deluxetable}{lcccccc}
\tabletypesize{\scriptsize}
\tablecaption{Observed \aco\ line parameters. 
\label{tab-03}}
\tablehead{
Source          & $z$                 & $\nu_{\rm obs}$ & $S_{\nu}$\tablenotemark{a} & $\Delta V_{\rm FWHM}$ & $I$ & Telescope \\
                &                     & [GHz]           & [mJy]                      & [\kms]                & [Jy \kms]        
}
\startdata
BR\,1202-0725   & 4.6932 $\pm$ 0.0004 & 20.2450         & 0.36 $\pm$ 0.03 & 329 $\pm$ 36          & 0.124 $\pm$ 0.012 & GBT \\
                & 4.6956 $\pm$ 0.0012 & 20.2388         & 0.40 $\pm$ 0.11 & 522 $\pm$ 146         & 0.22 $\pm$ 0.06 & Effelsberg \\
                & 4.6949 $\pm$ 0.0003 & 20.2411         & 0.34 $\pm$ 0.03 & 333 $\pm$ 30          & 0.120 $\pm$ 0.010 & (combined) \\
\tableline
PSS\,J2322+1944 & 4.1173 $\pm$ 0.0003 & 22.5258         & 0.77 $\pm$ 0.07 & 202 $\pm$ 17          & 0.165 $\pm$ 0.014 & GBT \\
                & 4.1184 $\pm$ 0.0008 & 22.5211         & 1.29 $\pm$ 0.26 & 184 $\pm$ 46          & 0.25 $\pm$ 0.06 & Effelsberg \\
                & 4.1179 $\pm$ 0.0002 & 22.5231         & 0.77 $\pm$ 0.07 & 190 $\pm$ 14          & 0.155 $\pm$ 0.013 & (combined) \\
\tableline
APM\,08279+5255 & 3.9122 $\pm$ 0.0007 & 23.4663         & 0.26 $\pm$ 0.04 & 556 $\pm$ 55          & 0.152 $\pm$ 0.020 & GBT \\
\vspace{-3mm}
\enddata
\tablenotetext{a}{A $T_{\rm A}^{\star}$(K)/$S$(Jy) conversion factor of 1.5 was assumed for the GBT.} 
\tablecomments{
All given uncertainties are formal (correlated) errors from the Gaussian fit. The $z$ error was derived from a full 3 parameter fit. 
In the error estimates for $S_{\nu}$ and $\Delta V_{\rm FWHM}$, the correlation with the error in $z$ is neglected, as the relative 
uncertainty in $z$ is by $\sim$3 orders of magnitudes less as that of the other two quantities. The error estimate for $I$ takes the correlation 
of the errors of $S_{\nu}$ and $\Delta V_{\rm FWHM}$ into account. 
}
\end{deluxetable}


\begin{deluxetable}{lcccccccc}
\tabletypesize{\scriptsize}
\tablecaption{Luminosities, gas masses and Star--formation rates (SFRs).\label{tab-04}}
\tablehead{
Source          & $D_{\rm L}$ & $\mu_{\rm L}^{\rm CO}$ & $L_{\rm FIR}$\tablenotemark{a} & $L'_{\rm CO(1-0)}$\tablenotemark{a} & $L_{\rm FIR}$/$L'_{\rm CO(1-0)}$ & $M_{\rm gas}({\rm H_2})$\tablenotemark{b} & $SFR$\tablenotemark{c} \\
                & [Gpc]       &     & [10$^{12}\,$L$_{\odot}$]       & [10$^{10}\,$K \kms pc$^2$] & [L$_{\odot}$/K \kms pc$^2$]        & [10$^{10}$\,M$_{\odot}$] & [10$^3$\,M$_{\odot}$\,yr$^{-1}$]%
}
\startdata
BR\,1202-0725   & 44.2        & 1\tablenotemark{d}   & 60\tablenotemark{g}/60         & 10.1/10.1                 &  596                             & 8.1                     & 9.0 \\
PSS\,J2322+1944 & 37.8        & 2.5\tablenotemark{e} & 23\tablenotemark{h}/9.2        & 10.5/4.2                  &  219                             & 3.4                     & 1.4 \\
APM\,08279+5255 & 35.6        & 7\tablenotemark{f}   & 200\tablenotemark{i}/28.6      &  9.6/1.4                  &  2090                            & 1.1                     & 4.3 \\
\vspace{-2mm}
\enddata
\tablenotetext{a}{Apparent luminosities (not corrected for lensing)/intrinsic luminosities (lensing--corrected).}
\tablenotetext{b}{Assuming a conversion factor of $\alpha = 0.8$\,M$_{\odot}\,$/K \kms pc$^2$ from $L'_{\rm CO(1-0)}$ to $M_{\rm gas}({\rm H_2})$ as appropriate for ULIRGs (see Downes \& Solomon 1998).}
\tablenotetext{c}{Assuming a Schmidt--Kennicutt law (Kennicutt 1998a,b): $SFR$[M$_{\odot}$\,yr$^{-1}$]$ = 1.5 \times 10^{-10}\,L_{\rm FIR}$[L$_{\odot}$], 
i.e., $\delta_{\rm MF} \delta_{\rm SB} = 1.5$ following the notation of Omont et al.\ (2001): $\delta_{\rm MF}$ describes the 
dependence on the mass function of the stellar population, $\delta_{\rm SB}$ gives the fraction of $L_{\rm FIR}$ that is actually 
powered by the starburst and not the AGN.}
\tablenotetext{d}{Carilli et al.\ (2002b);}
\tablenotetext{e}{Carilli et al.\ (2003);}
\tablenotetext{f}{${}$Lewis et al.\ (2002);}
\tablenotetext{g}{Carilli et al.\ (2005);}
\tablenotetext{h}{Cox et al.\ (2002);}
\tablenotetext{i}{Beelen et al.\ (2006).}
\end{deluxetable}


\begin{deluxetable}{lccccccl}
\tabletypesize{\scriptsize}
\tablecaption{
  CO detections in the quasar hosts of our targets in the literature. 
\label{tab-01}}
\tablehead{
Source                &  & $z$    & Transition & $S_{\nu}$       & $I$             & $\Delta V_{\rm FWHM}$& ref. \\
                      &  &        &            & [mJy]           & [Jy \kms]       & [\kms]               & 
}
\startdata
BR\,1202-0725         &N & 4.692  & 2--1       & 0.44 $\pm$ 0.07 & 0.26 $\pm$ 0.05 &                      & 1 \\
                      &S & 4.695  & 2--1       & 0.77 $\pm$ 0.10 & 0.23 $\pm$ 0.04 &                      & 1 \\
                      &NS&        & 4--3       & $\sim$5.1       & 1.50 $\pm$ 0.3  & 280 $\pm$ 30         & 2 \\
                      &N & 4.6916 & 5--4       & $\sim$3.5       & 1.3 $\pm$ 0.2   & 350                  & 2 \\
                      &S & 4.6947 & 5--4       & $\sim$5.5       & 1.1 $\pm$ 0.3   & 190                  & 2 \\
                      &NS& 4.695  & 5--4       & 9.3 $\pm$ 2.1   & 2.7 $\pm$ 0.41  & 220 $\pm$ 74         & 3 \\
                      &NS& 4.6915 & 7--6       & $\sim$10.6      & 3.1 $\pm$ 0.86  & $\sim$275            & 2 \\
PSS\,J2322+1944       &  & 4.1192 & 1--0       & 0.89 $\pm$ 0.22 & 0.19 $\pm$ 0.08 & 200 $\pm$ 70         & 4 \\
                      &  & 4.1192 & 2--1       & 2.70 $\pm$ 0.24 & 0.92 $\pm$ 0.30 & 280 $\pm$ 42         & 4,5 \\
                      &  & 4.1199 & 4--3       & 10.5            & 4.21 $\pm$ 0.40 & 375 $\pm$ 41         & 6 \\
                      &  & 4.1199 & 5--4       & 12              & 3.74 $\pm$ 0.56 & 273 $\pm$ 50         & 6 \\
                      &  &        & 6--5       &                 &                 &                      & 7 \\
                      &  &        & 7--6       &                 &                 &                      & 7 \\
APM\,08279+5255       &  & 3.9    & 1--0       &                 & 0.150 $\pm$ 0.045\tablenotemark{a}; 0.22 $\pm$ 0.05 & 575 & 8,9,10 \\
                      &  & 3.9    & 2--1       &                 & $\sim$0.81\tablenotemark{b} &                      & 8 \\
                      &  & 3.9114 & 4--3       & 7.4 $\pm$ 1.0   & 3.7 $\pm$ 0.5   & 480 $\pm$ 35         & 11 \\
                      &  &        & 6--5       &                 &                 &                      & 7 \\
                      &  & 3.9109 & 9--8       & 17.9 $\pm$ 1.4  & 9.1 $\pm$ 0.8   &                      & 11 \\
                      &  &        & 10--9      &                 &                 &                      & 7 \\
                      &  &        & 11--10     &                 &                 &                      & 7 \\
\vspace{-2mm}
\enddata
\tablenotetext{a}{Derived for central $\sim 1''$ only.}
\tablenotetext{b}{Assuming a velocity--averaged brightness temperature ratio between \bco\ and \aco\ of 1.35 $\pm$ 0.55 (Papadopoulos et al.\ 2001).}
\tablerefs{
[1] Carilli et al.\ (2002b); [2] Omont et al.\ (1996); [3] Ohta et al.\ (1996); [4] Carilli et al.\ (2002a); 
[5] Carilli et al.\ (2003); [6] Cox et al.\ (2002); [7] Weiss et al.\ (2006), in prep.; [8] Papadopoulos et al.\ (2001); 
[9] Lewis et al.\ (2002); [10] Riechers et al.\ (2006), in prep.; [11] Downes et al.\ (1999)}
\tablecomments{For BR\,1202-0725, N and S indicate the northern and southern component, NS indicates an integrated measurement over both components.}
\end{deluxetable}


\begin{thebibliography}{}
\bibitem[1992]{all92} Alloin, D., Barvainis, R., Gordon, M.~A., \&
  Antonucci, R.~R.~J.\ 1992, A\&A, 265, 429
\bibitem[2006]{bee06} Beelen, A., Cox, P., Benford, D.~J., et al.\
  2006, ApJ, 642, 694
\bibitem[2003]{ber03} Bertoldi, F., Cox, P., Neri, R., et al.\ 2003,
  A\&A, 409, L47
\bibitem[2002a]{car02a} Carilli, C.~L., Cox, P., Bertoldi, F., et al.\
  2002a, ApJ, 575, 145
\bibitem[2002b]{car02b} Carilli, C.~L., Kohno, K., Kawabe, R., et al.\
  2002b, AJ, 123, 1838
\bibitem[2003]{car03} Carilli, C.~L., Lewis, G.~F., Djorgovski, S.~G.,
  et al.\ 2003, Science, 300, 773
\bibitem[2005]{car05} Carilli, C.~L., Solomon, P., Vanden Bout, P., et
  al.\ 2005, ApJ, 618, 586
\bibitem[2002]{cox02} Cox, P., Omont, A., Djorgovski, S. G., et al.\
  2002, A\&A, 387, 406
\bibitem[1998]{dow98} Downes, D., \& Solomon, P.~M. 1998, ApJ, 507,
  615
\bibitem[1999]{dow99} Downes, D., Neri, R., Wiklind, T., et al.\ 1999,
  ApJ, 513, L1
\bibitem[2000]{ega00} Egami, E., Neugebauer, G., Soifer, B.~T., et
  al.\ 2000, ApJ, 535, 561
\bibitem[2001]{eva01} Evans, A.~S., Frayer, D.~T., Surace, J.~A., \&
  Sanders, D.~B.\ 2001, AJ, 121, 1893
\bibitem[2001]{fan01} Fan, X., Narayanan, V.~K., Lupton, R.~H., et
  al.\ 2001, AJ, 122, 2833
\bibitem[1999]{fix99} Fixsen, D.~J., Bennett, C.~L., \& Mather, J.~C.\
  1999, ApJ, 526, 207
\bibitem[2001]{flo01} Flower, D.~R., \& Pineau des Forets, G. 2001,
  MNRAS, 323, 672
\bibitem[2004]{gao04} Gao, Y., \& Solomon, P.~M.\ 2004, ApJ, 606, 271
\bibitem[2004]{gre04} Greve, T.~R., Ivison, R.~J., \& Papadopoulos,
  P.~P. 2004, A\&A, 419, 99
\bibitem[1999]{gui99} Guilloteau, S., Omont, A., Cox, P., et al.\
  1999, A\&A 349, 363
\bibitem[1996]{hu96} Hu, E.~M., McMahon, R.~G., \& Egami, E.\ 1996,
  ApJ, 459, L53
\bibitem[2002]{hu02} Hu, E.~M., Cowie, L.~L., McMahon, R. G., et al.\
  2002, ApJ, 568, L75
\bibitem[1999]{iba99} Ibata, R.~A., Lewis, G.~F., Irwin, M.~J., et
  al.\ 1999, AJ, 118, 1922
\bibitem[1998]{ken98b} Kennicutt, R.~C. 1998a, ApJ, 498, 541
\bibitem[1998]{ken98b} Kennicutt, R.~C. 1998b, ARA\&A, 36, 189
\bibitem[2003]{kod03} Kodaira, K., Taniguchi, Y., Kashikawa, N., et
  al.\ 2003, PASJ 55, L17
\bibitem[2004]{kla04} Klamer, I.~J., Ekers, R.~D., Sadler, E.~M., \&
  Hunstead, R.~W. 2004, ApJ, 612, L97
\bibitem[2005]{kla05} Klamer, I.~J., Ekers, R.~D., Sadler, E.~M., et
  al.\ 2005, ApJ, 621, L1
\bibitem[1998]{led98} Ledoux, C., Theodore, B., Petitjean, P., et al.\
  1998, A\&A, 339, L77
\bibitem[1998]{lew98} Lewis, G.~F., Chapman, S.~C., Ibata, R.~A., et
  al.\ 1998, ApJ, 505, L1
\bibitem[2002]{lew02} Lewis, G.~F., Carilli, C., Papadopoulos, P., \&
  Ivison, R.~J. 2002, MNRAS, 330, L15
\bibitem[1988]{moo88} Mooney, T.~J., \& Solomon, P.~M.\ 1988, ApJ,
  334, L51
\bibitem[1996]{oht96} Ohta, K., Yamada, T., Nakanishi, K., et al.\
  1996, Nature, 382, 426
\bibitem[1996]{omo96} Omont, A., Petitjean, P., Guilloteau, S., et
  al.\ 1996, Nature, 382, 428
\bibitem[2001]{omo01} Omont, A., Cox, P., Bertoldi, F., et al.\ 2001,
  A\&A, 374, 371
\bibitem[1994]{ott94} Ott, M., Witzel, A., Quirrenbach, A., et al.\
  1994, A\&A, 284, 331
\bibitem[2001]{pap01} Papadopoulos, P., Ivison, R., Carilli, C.~L., \&
  Lewis, G. 2001, Nature, 409, 58
\bibitem[2004]{pet04} Pety, J., Beelen, A., Cox, P., et al.\ 2004,
  A\&A, 428, L21
\bibitem[2001]{rho01} Rhoads, J.~E., \& Malhotra, S.\ 2001, ApJ, 563,
  L5
\bibitem[2000]{rr2000} Rowan-Robinson, M.\ 2000, MNRAS, 316, 885
\bibitem[1994]{san94} Sandage, A.\ 1994, ApJ, 430, 1
\bibitem[1991]{san91} Sanders, D.~B., Scoville, N.~Z., \& Soifer,
  B.~T.\ 1991, ApJ, 370, 158
\bibitem[2003]{sco03} Scoville, N.~Z., Frayer, D.~T., Schinnerer, E.,
  \& Christopher, M.\ 2003, ApJ, 585, L105
\bibitem[1992]{sol92} Solomon, P.~M., Radford, S.~J.~E., \& Downes,
  D.\ 1992, Nature, 356, 318
\bibitem[1997]{sol97} Solomon, P.~M., Downes, D., Radford, S.~J.~E.,
  \& Barrett, J.~W.\ 1997, ApJ, 478, 144
\bibitem[2003]{sol03} Solomon, P.~M., Vanden Bout, P.~A., Carilli,
  C.~L., \& Guelin, M.\ 2003, Nature 426, 636
\bibitem[2005]{sv05} Solomon, P.~M., \& Vanden Bout, P.~A. 2005,
  ARA\&A, 43, 677
\bibitem[2003]{spe03} Spergel, D.~N., Verde, L., Peiris, H.~V., et
  al.\ 2003, ApJS, 148, 175
\bibitem[2005]{tan05} Taniguchi, Y., Ajiki, M., Nagao, T., et al.\
  2005, PASJ, 57, 165
\bibitem[2004]{vdb04} Vanden Bout, P.~A., Solomon, P.~M., \&
  Maddalena, R.~J.\ 2004, ApJ, 614, L97
\bibitem[2003]{wal03} Walter, F., Bertoldi, F., Carilli, C.~L., et
  al.\ 2003, Nature, 424, 406
\bibitem[2004]{wal04} Walter, F., Carilli, C., Bertoldi, F., et al.\
  2004, ApJ, 615, L17
\bibitem[2001]{wei01} Weiss, A., Neininger, N., H\"uttemeister, S., \&
  Klein, U.\ 2001, A\&A, 365, 271
\bibitem[2003]{wei03} Weiss, A., Henkel, C., Downes, D., \& Walter,
  F.\ 2003, A\&A, 409, L41
\bibitem[2005]{wei05c} Weiss, A., Walter, F., \& Scoville, N.~Z.\
  2005, A\&A, 438, 533
\bibitem[2000]{yun00} Yun, M.~S., Carilli, C.~L., Kawabe, R., et al.\
  2000, ApJ, 528, 171
\bibitem[2004]{yun04} Yun, M.~S., Reddy, N.~A., Scoville, N.~Z., et
  al.\ 2004, ApJ, 601, 723

\end{thebibliography}
\end{document}